\documentstyle[12pt,epsfig]{article}
\def\bge{\begin{equation}}
\def\ene{\end{equation}}
\def\bg{\begin{eqnarray}}
\def\en{\end{eqnarray}}
\def\nn{\nonumber}

\begin{document}
\renewcommand{\thefootnote}{\fnsymbol{footnote}}
\renewcommand{\thesection}{\Roman{section}}
\begin{flushright}
ADP-96-40/T236 
\end{flushright}
%
%
\begin{center}
\begin{LARGE}
Variation of hadron masses in finite nuclei
\end{LARGE}
\end{center}
\vspace{0.5cm}
\begin{center}
\begin{large}
K.~Saito\footnote{ksaito@nucl.phys.tohoku.ac.jp} \\
Physics Division, Tohoku College of Pharmacy \\ Sendai 981, Japan \\
K.~Tsushima\footnote{ktsushim@physics.adelaide.edu.au} and 
A.~W.~Thomas\footnote{athomas@physics.adelaide.edu.au} \\
Department of Physics and Mathematical Physics \\
University of Adelaide, South Australia, 5005, Australia
\end{large}
\end{center}
\vspace{0.5cm}
\begin{abstract}
The quark-meson coupling model, based on a mean-field description of 
non-overlapping nucleon bags bound by the self-consistent exchange of 
$\sigma$, $\omega$ and $\rho$ mesons, is extended to investigate the 
change of hadron properties in finite nuclei.  Relativistic Hartree equations 
for spherical nuclei have been derived from a relativistic quark model of 
the structure of bound nucleons and mesons.  
Using this unified, self-consistent description of both infinite nuclear 
matter and finite nuclei, we investigate the properties of some closed-shell 
nuclei, and study the changes in the hadron masses of 
the non-strange vector mesons, the hyperons and the nucleon in those nuclei.  
We find a new, simple scaling relation for the changes of the hadron masses, 
which can be described in terms of the number of non-strange quarks in the 
hadron and the value of the scalar mean-field in a nucleus. 
\end{abstract}
PACS numbers: 12.39.Ba, 21.60.-n, 21.90.+f, 24.85.+p 
%
%
\newpage
\section{INTRODUCTION}

One of the most exciting topics in nuclear physics is the 
study of the variation of
hadron properties as the nuclear environment 
changes. In particular, the medium modification of the light vector-mesons 
is receiving a lot of attention, both theoretically and experimentally.  
Recent experiments from the HELIOS-3~\cite{helios} and the CERES~\cite{ceres} 
collaborations at the SPS/CERN energies have shown that there exists a large 
excess of the $e^+ e^-$ pairs in central S + Au collisions.  Those 
experimental results may give a hint of some change of hadron 
properties in nuclei~\cite{li}.  
Forthcoming, ultra-relativistic heavy-ion experiments (eg. at RHIC) 
are also expected to give significant information on the strong interaction 
(QCD), through the detection of changes in hadronic properties (for a review, 
see Ref.\cite{review}). 

Theoretically, lattice QCD simulations may eventually give the most 
reliable information on the density and/or temperature dependence of hadron 
properties in matter.  However, current simulations have been performed only 
for finite temperature systems with zero baryon density~\cite{lattice}. 
Therefore, many authors have studied the hadron masses in matter using 
effective theories: 
the vector dominance model~\cite{asa}, 
QCD sum rules~\cite{hatas} and the Walecka model~\cite{sai1,suzuki,will,hat}, 
and have reported that the mass decreases in the nuclear medium (see also 
Ref.\cite{hat0}).  

In the approach based on QCD sum rules, the reduction of 
the mass is mainly due to the four-quark condensates and one of the 
twist-2 condensates. However, it has been suggested that there may 
be considerable, intrinsic  
uncertainty in the standard assumptions underlying the QCD sum-rule 
analyses~\cite{griegel}.  In hadronic models, like Quantum hadrodynamics 
(QHD)~\cite{walecka}, the on-shell properties of the scalar ($\sigma$) and 
vector ($\omega$) meson with vacuum polarization were first studied by Saito, 
Maruyama and Soutome~\cite{sai1}, and later by many 
authors~\cite{suzuki,will,hat}. (Good physical arguments concerning
the $\omega$ meson in medium were found in Ref.\cite{will}.) 
The main reason for the reduction in masses in QHD is the polarization of 
the Dirac sea, where the {\em anti-nucleons} in matter play a 
crucial role.  From the point of view of the quark model, however, the strong 
excitation of $N$-${\bar N}$ pairs in medium is difficult to 
understand~\cite{sai2}. 

Recently Guichon, Saito, Rodionov and Thomas~\cite{guichon1} have developed 
an entirely different model for both nuclear matter and finite nuclei, 
in which quarks in non-overlapping nucleon bags interact 
{\em self-consistently} with (structureless) scalar ($\sigma$) and 
vector ($\omega$ and $\rho$) mesons in the mean-field approximation (MFA) 
-- the quark-meson coupling (QMC) model.  
(The original idea was proposed by Guichon in 1988~\cite{guichon2}. 
Several interesting applications to the properties of nuclear matter and 
finite nuclei are also given in a series of papers by Saito and 
Thomas~\cite{sai2,sai3,sai35,sai4}.)  
This model was recently used to calculate detailed properties of static, 
closed shell nuclei from $^{16}$O to $^{208}$Pb, where it was shown that 
the model can reproduce fairly well the observed charge density 
distributions, neutron density distributions etc.~\cite{sai5}.  
Blunden and Miller~\cite{blun} have also considered a model for finite nuclei 
along this line.  

To investigate the properties of hadrons, particularly the changes in their 
masses in nuclear medium, one must also consider the structure 
of the mesons, as well as the nucleon.  Saito and Thomas~\cite{sai4} have 
studied variations of hadron masses and matter properties in {\em infinite} 
nuclear matter, in which the vector-mesons are also described by bags, but 
the scalar-meson mass is kept constant, and have shown the decrease of 
the hadron mass.  Now it would be most desirable to extend 
this picture to {\em finite\/} nuclei to study the changes of  
hadron properties in the medium, {\em quantitatively}.  

Our main aim in this paper is to give an effective Lagrangian density for 
finite nuclei, in which the structure effects of the mesons ($\sigma$, 
$\omega$ and $\rho$) as well as 
the nucleon are involved, and to study quantitative changes in the hadron 
(including the hyperon) masses by solving relativistic Hartree equations 
for spherical nuclei derived 
from the Lagrangian density. (Using this model, we also calculate some 
static properties of closed-shell nuclei.) 
In the present model the change in the 
hadron mass can be described by a simple formula, which is expressed in 
terms of the number of non-strange quarks 
and the value of the scalar mean-field (see also Ref.\cite{sai4}).  
This is accurate over a wide range of nuclear density.  We then find a new, 
simple scaling relation for the changes of hadron masses in the medium: 
\bge
\frac{\delta m_{\omega, \rho}^{\star}}{\delta M_N^{\star}} \sim 
\frac{\delta M_{\Lambda, \Sigma}^{\star}}{\delta M_N^{\star}} \sim 
\frac{2}{3}, \ \ \ 
\frac{\delta M_{\Xi}^{\star}}{\delta M_N^{\star}} \sim \frac{1}{3}, 
\ \ \ \mbox{etc.}
\ene
where $\delta M_i^{\star} \equiv M_i - M_i^{\star}$, with the effective 
hadron mass, $M_i^{\star}$ ($i= N, \omega, \rho, \cdots$).   

An outline of the paper is as follows.  In Sec.~\ref{sec:qmc}, the idea of 
the QMC model is first reviewed.  Then, the model is extended to include 
the effect of meson structure.  In Sec.~\ref{sec:numerical}, 
parameters in the model are first determined to reproduce the properties of 
infinite nuclear matter, and the hadron masses in the medium are then 
discussed.  A new scaling relationship among them is also derived.  The 
static properties of several closed-shell nuclei are studied in 
subsection~\ref{subsec:finite}, where we also show the changes of the masses 
of the nucleon, the mesons ($\sigma$, $\omega$ and $\rho$) and the hyperons 
($\Lambda$, $\Sigma$ and $\Xi$) in $^{40}$Ca and $^{208}$Pb.  The last 
section gives our conclusions. 

\section{THE QUARK-MESON COUPLING MODEL}
\label{sec:qmc}
\subsection{Effect of nucleon structure in finite nuclei}
\label{subsec:nucleon}

Let us suppose that a free nucleon (at the origin) consists of three light 
(u and d) quarks under a (Lorentz scalar) confinement potential, $V_c$.  
Then, the Dirac equation for the quark field, $\psi_q$, is given by 
\bge
[ i\gamma\cdot\partial - m_q - V_c(r) ] \psi_q(r) = 0 , 
\label{dirac1}
\ene
where $m_q$ is the bare quark mass.  

Next we consider how Eq.(\ref{dirac1}) is modified when the nucleon is bound 
in static, uniformly distributed (iso-symmetric) nuclear matter.  
In the QMC model~\cite{guichon2} it is assumed that 
each quark feels scalar, $V_s^q$, and vector, $V_v^q$, potentials, which are 
generated by the
surrounding nucleons, as well as the confinement potential (see 
also Ref.\cite{blun}). 
Since the typical distance between two nucleons around normal nuclear density 
($\rho_0 = 0.15$ fm$^{-3}$) is surely larger than the typical size of the 
nucleon (the radius $R_N$ is about 0.8 fm), the interaction (except for 
the short-range part) between the nucleons 
should be colour singlet; e.g., a meson-exchange potential.  Therefore, this 
assumption seems appropriate when the baryon density, $\rho_B$, is not high.  
If we use the mean-field approximation for the meson fields, 
Eq.(\ref{dirac1}) may be rewritten as 
\bge
[ i\gamma\cdot\partial - (m_q - V_s^q) - V_c({\vec r}) 
 - \gamma_0 V_v^q ] \psi_q({\vec r}) = 0 . 
\label{dirac2}
\ene
The potentials generated by the medium are constants because the matter 
distributes uniformly. As the nucleon is static, the time-derivative 
operator in the Dirac equation can be 
replaced by the quark energy, $-i \epsilon_q$.  
By analogy with the procedure applied to the nucleon
in QHD~\cite{walecka}, if we introduce the 
effective quark mass by $m_q^{\star} = m_q - V_s^q$, the Dirac equation, 
Eq.(\ref{dirac2}), can be rewritten in the same form as that in free space, 
with the mass $m_q^{\star}$ and the energy $\epsilon_q - V_v^q$, instead of 
$m_q$ and $\epsilon_q$. 
In other words, the vector interaction has {\em no effect 
on the nucleon structure} except for an overall phase in the quark wave 
function, which gives a shift in the nucleon energy.  This fact 
{\em does not\/} depend on how to choose the confinement potential, $V_c$.  
Then, the nucleon energy (at rest), $E_N$, in the medium is~\cite{sai35} 
\bge
E_N = M_N^{\star}(V_s^q) + 3V_v^q , 
\label{efmas}
\ene
where the effective nucleon mass, $M_N^{\star}$, depends on {\em only the 
scalar potential\/} in the medium.  

Now we extend this idea to finite nuclei. 
The solution of the general problem of a composite, quantum particle 
moving in background scalar and vector fields that vary with position is 
extremely difficult.  One has, however, a chance to solve the particular 
problem of interest to us, namely light quarks confined in a nucleon which is 
itself bound in a finite nucleus, only because the nucleon motion is 
relatively slow and the quarks highly relativistic~\cite{guichon1}. Thus the 
Born-Oppenheimer approximation, in which the nucleon internal 
structure has time to adjust to the local fields, is naturally suited to 
the problem. It is relatively easy to establish that the method should 
be reliable at the level of a few percent~\cite{guichon1}. 

Even within the Born-Oppenheimer approximation, the nuclear surface 
gives rise to external fields that may vary appreciably 
across the finite size of the nucleon.  Our 
approach in Ref.\cite{guichon1} was to start with a classical nucleon 
and to allow its internal structure 
to adjust to minimise the energy of three quarks in the ground-state of 
a system under constant scalar and vector fields, 
with values equal to those at the centre of the nucleon. 
In Ref.\cite{guichon1}, the MIT bag model was used to describe the nucleon 
structure.  Blunden and Miller have also examined a relativistic oscillator 
model as an alternative model~\cite{blun}.  
Of course, the major problem with 
the MIT bag (as with many other relativistic models of nucleon structure) 
is that it is difficult to boost. We therefore solve the bag 
equations in the instantaneous rest frame (IRF) of the nucleon -- using a
standard Lorentz transformation to find the energy and momentum of the
classical nucleon bag in the nuclear rest frame. 
Having solved the problem using the meson fields at the centre of 
the $\lq\lq$nucleon'' (which is a quasi-particle with nucleon quantum 
numbers), 
one can use perturbation theory to correct for the variation of the 
scalar and vector fields across the nucleon bag. In first order perturbation 
theory only the spatial components of the vector potential 
give a non-vanishing contribution. (Note that, although in the nuclear 
rest frame only the time component of the vector field is non-zero, 
in the IRF of the nucleon there are also non-vanishing spatial components.) 
This extra term is a correction to the spin-orbit force. 

As shown in Refs.\cite{guichon1,sai5}, the basic result in the QMC model 
is that, in the scalar ($\sigma$) and vector ($\omega$) 
meson fields, the nucleon behaves essentially as a point-like 
particle with an effective mass 
$M_N^{\star}$, which depends on the position through only the $\sigma$ 
field, moving in a vector potential generated by the $\omega$ meson, as 
mentioned near Eq.(\ref{efmas}).  Although we discussed the QMC model using 
the specific model, namely the bag model, in Ref.\cite{guichon1,sai5}, 
{\em the qualitative features 
we found are correct in any model\/} in which the nucleon contains 
{\em relativistic quarks\/} and the (middle- and long-range) 
{\em attractive\/} 
and (short-range) {\em repulsive\/} N-N forces have {\em Lorentz-scalar\/} and 
{\em vector characters}, respectively.  

Let us suppose that the scalar and vector potentials in Eq.(\ref{dirac2}) are 
mediated by the $\sigma$ and $\omega$ mesons, and introduce their 
mean-field values, which now depend on position ${\vec r}$, 
by $V_s^q({\vec r}) = g_{\sigma}^q \sigma({\vec r})$ and 
$V_v^q({\vec r}) = g_{\omega}^q \omega({\vec r})$, respectively, where 
$g_{\sigma}^q$ ($g_{\omega}^q$) is the coupling constant of the quark-$\sigma$ 
($\omega$) meson.  Furthermore, we shall add the isovector, vector meson, 
$\rho$, and the Coulomb field, $A({\vec r})$, to describe finite nuclei 
realistically~\cite{guichon1,sai5}.  
Then, the effective Lagrangian density for finite nuclei, involving the quark 
degrees of freedom in the nucleon and the (structureless) meson fields, 
in MFA would be given by~\cite{sai5}
\bg
{\cal L}_{QMC-I}&=& \overline{\psi} [i \gamma \cdot \partial 
- M_N^{\star}(\sigma({\vec r})) 
- g_\omega \omega({\vec r}) \gamma_0 \nn \\
&-& g_\rho \frac{\tau^N_3}{2} b({\vec r}) \gamma_0 
- \frac{e}{2} (1+\tau^N_3) A({\vec r}) \gamma_0 ] \psi \nn \\
&-& \frac{1}{2}[ (\nabla \sigma({\vec r}))^2 + 
m_{\sigma}^2 \sigma({\vec r})^2 ] 
+ \frac{1}{2}[ (\nabla \omega({\vec r}))^2 + m_{\omega}^2 
\omega({\vec r})^2 ] \nn \\
&+& \frac{1}{2}[ (\nabla b({\vec r}))^2 + m_{\rho}^2 b({\vec r})^2 ] 
+ \frac{1}{2} (\nabla A({\vec r}))^2 , 
\label{qmclag}
\label{qmc-1}
\en
where $\psi({\vec r})$ and $b({\vec r})$ are respectively the 
nucleon and the $\rho$ (the time component in the third direction of 
isospin) fields. $m_\sigma$, $m_\omega$ and $m_{\rho}$ are respectively 
the (constant) masses of the $\sigma$, $\omega$ and $\rho$ mesons. 
$g_\omega$ and $g_{\rho}$ are respectively the $\omega$-N and $\rho$-N 
coupling constants, which are related to the corresponding quark-$\omega$, 
$g_\omega^q$, and quark-$\rho$, $g_\rho^q$, coupling constants as 
$g_\omega = 3 g_\omega^q$ and $g_\rho = g_\rho^q$~\cite{guichon1,sai5}. 
We call this model the QMC-I model.  
If we define the field-dependent $\sigma$-N coupling 
constant, $g_\sigma(\sigma)$, by
\bge
M_N^{\star}(\sigma({\vec r})) \equiv M_N - g_\sigma(\sigma({\vec r})) 
\sigma({\vec r}) , \label{coup}
\ene
where $M_N$ is the free nucleon mass, it is easy to compare with 
QHD~\cite{walecka}.  $g_\sigma(\sigma)$ will be discussed further below. 

The difference between QMC-I  
and QHD lies only in the coupling constant $g_\sigma$, which
depends on the scalar field in QMC-I while it is constant in QHD.  
(The relationship between QMC and QHD has been already clarified in 
Ref.\cite{sai35}.  See also Ref.\cite{jin1}.)  
However, this difference leads to 
a lot of favorable results, notably the nuclear compressibility, 
\cite{guichon1,sai3,sai35,sai5}.  Detailed calculated properties of both 
infinite nuclear matter and finite nuclei can be found in 
Refs.\cite{guichon1,sai5}.  

Here we consider the nucleon mass in matter further.  The nucleon mass is a 
function of the scalar field.  Because the scalar field is small 
at low density the nucleon mass can be expanded in terms of $\sigma$ as 
\bge
M_N^{\star} = M_N + \left( \frac{\partial M_N^{\star}}{\partial \sigma} 
\right)_{\sigma=0} \sigma + \frac{1}{2} \left( \frac{\partial^2 M_N^{\star}}
{\partial \sigma^2} \right)_{\sigma=0} \sigma^2 + \cdots . 
\label{nuclm}
\ene
In the QMC model the interaction Hamiltonian between the nucleon and the 
$\sigma$ field at the quark level is given by $H_{int} = - 3 g_{\sigma}^q 
\int d{\vec r} \ \overline{\psi}_q \sigma \psi_q$, and the derivative of 
$M_N^{\star}$ with respect to $\sigma$ is 
\bge
\left( \frac{\partial M_N^{\star}}{\partial \sigma} \right) 
= -3g_{\sigma}^q \int d{\vec r} \ \ {\overline \psi}_q \psi_q 
\equiv -3g_{\sigma}^q S_N(\sigma) . \label{deriv}
\ene
Here we have defined the quark-scalar density in the nucleon, $S_N(\sigma)$,
which is itself a function of the scalar field, by 
Eq.(\ref{deriv}). 
Because of a negative value of 
$\left( \frac{\partial M_N^{\star}}{\partial \sigma} \right)$, 
the nucleon mass decreases in matter at low density.  

Furthermore, we define the scalar-density ratio, $S_N(\sigma)/S_N(0)$, 
to be $C_N(\sigma)$ and the $\sigma$-N coupling constant at $\sigma = 0$  
to be 
$g_\sigma$ (i.e., $g_\sigma \equiv g_\sigma(\sigma=0)$):
\bge
C_N(\sigma) = S_N(\sigma)/S_N(0) \ \ \mbox{and} \ \ 
g_{\sigma} = 3g_{\sigma}^q S_N(0) . \label{cn}
\ene
Comparing with Eq.(\ref{coup}), we find that 
\bge
\left( \frac{\partial M_N^{\star}}{\partial \sigma} \right) 
= -g_{\sigma} C_N(\sigma) = - \frac{\partial}{\partial \sigma}
\left[ g_\sigma(\sigma) \sigma \right],
\label{deriv2}
\ene
and that the nucleon mass is 
\bge
M_N^{\star} = M_N - g_{\sigma} \sigma - \frac{1}{2} g_{\sigma} 
C_N^\prime(0) \sigma^2 + \cdots . 
\label{nuclm2}
\ene
In general, $C_N$ is a decreasing function because the quark in matter is 
more relativistic than in free space.  Thus, $C_N^\prime(0)$ takes a 
negative value. If the nucleon were structureless $C_N$ would not depend on 
the scalar field, that is, $C_N$ would be constant ($C_N=1$).  Therefore, 
only the first two terms in the right hand side of Eq.(\ref{nuclm2}) remain, 
which is exactly the same as the equation for the effective nucleon 
mass in QHD.  By taking the heavy-quark-mass limit in QMC we can reproduce 
the QHD results~\cite{sai35}.  

If the MIT bag model is adopted as the nucleon model, $S_N$ is explicitly 
given by~\cite{sai35,cmcor}
\bge
S_N(\sigma) = \frac{\Omega^{\star}/2 + m_q^{\star}R_N^{\star}
(\Omega^{\star}-1)} 
{\Omega^{\star}(\Omega^{\star}-1) + m_q^{\star}R_N^{\star}/2}, \label{sss}
\ene
where $\Omega^{\star} = \sqrt{x_N^{\star 2} + (R_N^{\star}m_q^{\star})^2}$ is 
the kinetic energy of the quark in units of $1/R_N^{\star}$ and 
$x_N^{\star}$ is the eigenvalue of the quark in the nucleon in matter.  
We denote the bag radius of the nucleon 
in free space (matter) by $R_N$ ($R_N^{\star}$).  
In actual numerical calculations we found that 
the scalar-density ratio, $C_N(\sigma)$, decreases linearly 
(to a very good approximation) with $g_{\sigma} 
\sigma$~\cite{guichon1,sai5}.  Then, it is very useful to have a simple 
parametrization for $C_N$: 
\bge
C_N(\sigma) = 1 - a_N \times (g_{\sigma} \sigma) , 
\label{paramCN}
\ene
with $g_{\sigma} \sigma$ in MeV (recall $g_{\sigma} = 
g_{\sigma}(\sigma=0)$) and $a_N \sim 9 \times 10^{-4}$ (MeV$^{-1}$) for 
$m_q$ = 5 MeV and $R_N$ = 0.8 fm.  
This is quite accurate up to $\sim 3\rho_0$. 

As a practical matter, it is easy to solve Eq.(\ref{deriv2}) for 
$g_\sigma(\sigma)$ in the 
case where $C(\sigma)$ is linear in $g_{\sigma} \sigma$, as in 
Eq.(\ref{paramCN}).  Then one finds 
\bge
M^{\star}_N = M_N - g_\sigma \left[ 1 - \frac{a_N}{2} (g_\sigma 
\sigma) \right] \sigma , 
\label{mstaR}
\ene
so that the effective $\sigma$-N coupling constant, $g_{\sigma}(\sigma)$, 
decreases at half the rate of $C_N(\sigma)$.  

\subsection{Effect of meson structure}
\label{subsec:meson}

In the previous section we have considered the effect of nucleon 
structure.  It is however true that the mesons are also built of quarks and 
anti-quarks, and that they may change their properties in matter.  

To incorporate the effect of meson structure in the QMC model, we 
suppose that the vector mesons are again described by a relativistic quark 
model with {\em common\/} scalar and vector mean-fields~\cite{sai4}, 
like the nucleon (see Eq.(\ref{dirac2})).  
Then, again the effective vector-meson mass in matter, 
$m_v^{\star} (v = \omega, \rho)$, depends on only the scalar mean-field.  

However, for the scalar ($\sigma$) meson it may not be easy to describe it 
by a simple quark model (like a bag) because it couples strongly 
to the pseudoscalar ($2 \pi$) channel, which requires a direct 
treatment of chiral symmetry in medium~\cite{hatkun}.  Since, according to the 
Nambu--Jona-Lasinio model~\cite{hatkun,bern} or the Walecka 
model~\cite{sai1}, one might expect the 
$\sigma$-meson mass in medium, $m_{\sigma}^{\star}$, to be less than the free 
one, we shall here parametrize it using a quadratic 
function of the scalar field: 
\bge
\left( \frac{m_{\sigma}^{\star}}{m_{\sigma}} \right) = 1 - a_{\sigma} 
(g_{\sigma} \sigma) + b_{\sigma} (g_{\sigma} \sigma)^2 , 
\label{sigmas}
\ene
with $g_{\sigma} \sigma$ in MeV, and we introduce two parameters, 
$a_{\sigma}$ (in MeV$^{-1}$) and $b_{\sigma}$ (in MeV$^{-2}$).  
(We will determine these parameters in the next section.)  

Using these effective meson masses, we can find a new Lagrangian 
density for finite nuclei, 
which involves the structure effects of not only the nucleons but also the 
mesons, in the MFA: 
\bg
{\cal L}_{QMC-II}&=& \overline{\psi} [i \gamma \cdot \partial 
- M_N^{\star} - g_\omega \omega({\vec r}) \gamma_0 
- g_\rho \frac{\tau^N_3}{2} b({\vec r}) \gamma_0 
- \frac{e}{2} (1+\tau^N_3) A({\vec r}) \gamma_0 ] \psi \nn \\
&-& \frac{1}{2}[ (\nabla \sigma({\vec r}))^2 + 
m_{\sigma}^{\star 2} \sigma({\vec r})^2 ] 
+ \frac{1}{2}[ (\nabla \omega({\vec r}))^2 + m_{\omega}^{\star 2} 
\omega({\vec r})^2 ] \nn \\
&+& \frac{1}{2}[ (\nabla b({\vec r}))^2 + m_{\rho}^{\star 2} b({\vec r})^2 ] 
+ \frac{1}{2} (\nabla A({\vec r}))^2 , 
\label{qmc-2}
\en
where the masses of the mesons and the nucleon depend on the scalar 
mean-fields.  We call this model QMC-II.  

At low density the vector-meson mass can be again expanded in the same 
way as in the nucleon case (Eq.(\ref{nuclm})): 
\bg
m_v^{\star} &=& m_v + \left( \frac{\partial m_v^{\star}}{\partial \sigma} 
\right)_{\sigma=0} \sigma + \frac{1}{2} \left( \frac{\partial^2 m_v^{\star}}
{\partial \sigma^2} \right)_{\sigma=0} \sigma^2 + \cdots , \nn \\
  &\simeq&  m_v - 2 g_{\sigma}^q S_v(0) \sigma - g_{\sigma}^q S_v^\prime(0) 
\sigma^2 , \nn \\
  &\equiv& m_v - \frac{2}{3} g_\sigma \Gamma_{v/N} \sigma 
  - \frac{1}{3} g_\sigma \Gamma_{v/N} C_v^\prime(0) \sigma^2 , 
\label{vmm}
\en
where $S_v(\sigma)$ is the quark-scalar density in the vector meson, 
\bge
\left( \frac{\partial m_v^{\star}}{\partial \sigma} \right) 
= - \frac{2}{3} g_\sigma \Gamma_{v/N} C_v(\sigma) , 
\label{deriv3}
\ene
and $C_v(\sigma) = S_v(\sigma)/S_v(0)$.  In Eqs.(\ref{vmm}) and 
(\ref{deriv3}), we introduce a correction factor, $\Gamma_{v/N}$, 
which is given by $S_v(0)/S_N(0)$, because the coupling 
constant, $g_\sigma$, is defined specifically for the nucleon
by Eq.(\ref{cn}).  

\section{NUMERICAL RESULTS}
\label{sec:numerical}

In this section we will show our numerical results using the Lagrangian 
density of the QMC-II model that is, including self-consistently the 
density dependence of the meson masses. 
We have studied the QMC-I model, and have 
already shown the calculated properties of finite 
nuclei in Refs.\cite{guichon1,sai5}.  

\subsection{Infinite nuclear matter}
\label{subsec:infinite}

For infinite nuclear matter we take the Fermi momenta for protons and 
neutrons to be $k_{F_i}$ ($i=p$ or $n$). This is 
defined by $\rho_i = k_{F_i}^3 / (3\pi^2)$, where $\rho_i$ is the 
density of protons or neutrons, and the total baryon density, 
$\rho_B$, is then given by $\rho_p + \rho_n$.  
Let the {\em constant\/} mean-field values for the $\sigma$, $\omega$ and 
$\rho$ fields be $\bar{\sigma}$, $\bar{\omega}$ and $\bar{b}$, 
respectively.  

>From the Lagrangian density Eq.(\ref{qmc-2}), the total energy per nucleon, 
$E_{tot}/A$, can be written (without the Coulomb force)
\bge
E_{tot}/A = \frac{2}{\rho_B (2\pi)^3}\sum_{i=p,n}\int^{k_{F_i}} 
d\vec{k} \sqrt{M_i^{\star 2} + \vec{k}^2} + \frac{m_{\sigma}^{\star 2}}
{2\rho_B}{\overline \sigma}^2 + \frac{g_{\omega}^2}
{2m_{\omega}^{\star 2}}\rho_B + \frac{g_{\rho}^2}{8m_{\rho}^{\star 2}
\rho_B} \rho_3^2 , \label{tote}
\ene
where the value of the $\omega$ field is now determined by baryon number 
conservation as ${\overline \omega}=g_{\omega}\rho_B / m_{\omega}^{\star 2}$, 
and the $\rho$-field value by the difference in proton and neutron 
densities, $\rho_3 = \rho_p - \rho_n$, as ${\overline b} = g_{\rho} \rho_3 / 
(2m_{\rho}^{\star 2})$~\cite{sai4}. 

On the other hand, the scalar mean-field is given by a self-consistency 
condition (SCC): 
\bg
{\overline \sigma} = &-& \frac{2}{(2\pi)^3 m_{\sigma}^{\star 2}} 
\left[ \sum_{i=p,n} \int^{k_{F_i}} d\vec{k} \frac{M_i^{\star}}
{\sqrt{M_i^{\star 2} + \vec{k}^2}} \left(\frac{\partial 
M_i^{\star}}{\partial {\overline \sigma}}\right) \right] \nn \\
 &+& \frac{g_\omega^2 \rho_B^2}{m_\omega^{\star 3} m_\sigma^{\star 2}} 
\left(\frac{\partial m_\omega^{\star}}{\partial {\overline \sigma}}\right) 
 + \frac{g_\rho^2 \rho_3^2}{4 m_\rho^{\star 3} m_\sigma^{\star 2}} 
\left(\frac{\partial m_\rho^{\star}}{\partial {\overline \sigma}}\right) 
 - \frac{{\overline \sigma}^2}{m_\sigma^\star} 
\left(\frac{\partial m_\sigma^{\star}}{\partial {\overline \sigma}}\right) .
\label{scc}
\en
Using Eqs.(\ref{deriv2}), (\ref{sigmas}) and (\ref{deriv3}), Eq.(\ref{scc}) 
can be rewritten 
\bg
{\overline \sigma} &=& \frac{2g_\sigma}{(2\pi)^3 m_{\sigma}^{\star 2}} 
\left[ \sum_{i=p,n} C_i({\overline \sigma}) 
\int^{k_{F_i}} d\vec{k} \frac{M_i^{\star}}
{\sqrt{M_i^{\star 2} + \vec{k}^2}} \right] 
 + g_\sigma \left( \frac{m_\sigma}{m_\sigma^\star} \right) 
 \left[ a_\sigma - 2 b_\sigma (g_\sigma {\overline \sigma} ) \right] 
{\overline \sigma}^2  \nn \\
 &-& \frac{2}{3} \left( \frac{g_\sigma}{m_\sigma^{\star 2}} \right) 
\left[ \frac{g_\omega^2 \rho_B^2}
{m_\omega^{\star 3}} \Gamma_{\omega/N} C_\omega({\overline 
\sigma}) + \frac{g_\rho^2 \rho_3^2}{4 m_\rho^{\star 3}} 
\Gamma_{\rho/N} C_\rho({\overline \sigma}) \right]. 
\label{scc2}
\en

Now we need a model for the structure of the hadrons involved.  We use the MIT 
bag model in static, spherical cavity approximation~\cite{MIT}.  
As in Ref.\cite{sai5}, the bag constant $B$ and the 
parameter $z_N$ (which accounts for the sum of the c.m. and 
gluon fluctuation corrections~\cite{guichon1}) in the familiar form of the 
MIT bag model Lagrangian are fixed to reproduce the free nucleon mass 
($M_N$ = 939 MeV) under the  
condition that the hadron mass be stationary under variation of
the free bag radius ($R_N$ in the case of the nucleon).  
Furthermore, to fit the free vector-meson masses, 
$m_{\omega}$ = 783 MeV and $m_{\rho}$ = 770 MeV, we introduce new 
$z$-parameters for them, $z_{\omega}$ and $z_{\rho}$.  
In the following we choose $R_N=0.8$ fm and the free quark mass 
$m_q$ = 5 MeV.  Variations of the quark mass and $R_N$ only lead to 
numerically small changes in the calculated results~\cite{sai5}.  
We then find that $B^{1/4}$ = 170.0 MeV, $z_N$ = 3.295, 
$z_\omega$ = 1.907 and $z_\rho$ = 1.857.  
Thus, $C_N$ is given by Eq.(\ref{sss}), and $C_v$ is given by a 
similar form, with the kinetic energy of quark and the bag radius for the 
vector meson.  We find that the bag model gives $\Gamma_{\omega, \rho/N}$ 
= 0.9996.  Therefore, we may discard those correction factors in practical 
calculations.  

Next we must choose the two parameters in the parametrization for the 
$\sigma$-meson mass in matter (see Eq.(\ref{sigmas})).  In this paper, we 
consider three parameter sets: 
(A) $a_\sigma = 3.0 \times 10^{-4}$ (MeV$^{-1}$) and 
$b_\sigma = 100 \times 10^{-8}$ (MeV$^{-2}$), 
(B) $a_\sigma = 5.0 \times 10^{-4}$ (MeV$^{-1}$) and 
$b_\sigma = 50 \times 10^{-8}$ (MeV$^{-2}$), 
(C) $a_\sigma = 7.5 \times 10^{-4}$ (MeV$^{-1}$) and 
$b_\sigma = 100 \times 10^{-8}$ (MeV$^{-2}$).  The parameter sets A, B and 
C give about 2\%, 7\% and 10\% decreases of the $\sigma$ mass 
at saturation density, respectively.
We will revisit this issue in the next subsection.

Now we are in a position to determine the coupling constants.  
$g_{\sigma}^2$ and $g_{\omega}^2$ are fixed to fit the binding energy 
($-15.7$ MeV) at the saturation density ($\rho_0 = 0.15$ fm$^{-3}$) for 
symmetric nuclear matter.  Furthermore, the $\rho$-meson coupling constant is 
used to reproduce the bulk symmetry energy, 35 MeV.  
We take $m_\sigma$ = 550 MeV. The coupling constants 
and some calculated properties for matter are listed in Table~\ref{ccc}.  
The last three columns show the relative changes (from their values at 
zero density) of the nucleon-bag radius 
($\delta R_N^{\star}/R_N$), the lowest eigenvalue ($\delta x_N^{\star}/x_N$) 
and the root-mean-square radius (rms radius) of the nucleon calculated using 
the quark wave function ($\delta r_q^{\star}/r_q$) at saturation density. 
\begin{table}[htbp]
\begin{center}
\caption{Coupling constants and calculated properties for 
symmetric nuclear matter at normal nuclear density ($m_q$ = 5 MeV, 
$R_N$ = 0.8 fm and $m_\sigma$ = 550 MeV).  
The effective nucleon mass, $M_N^{\star}$, and the nuclear 
compressibility, $K$, are quoted in MeV. 
The bottom row is for QHD.}
\label{ccc}
\begin{tabular}[t]{c|cccccccc}
\hline
type & $g_{\sigma}^2/4\pi$&$g_{\omega}^2/4\pi$&$g_\rho^2/4\pi$&
$M_N^{\star}$&$K$&$\delta R_N^{\star}/R_N$&$\delta x_N^{\star}/x_N$&$\delta 
r_q^{\star}/r_q$ \\
\hline
 A &  3.84 & 2.70 & 5.54 & 801 & 325 & $-0.01$ & $-0.11$ & 0.02 \\
 B &  3.94 & 3.17 & 5.27 & 781 & 382 & $-0.01$ & $-0.13$ & 0.02 \\
 C &  3.84 & 3.31 & 5.18 & 775 & 433 & $-0.02$ & $-0.14$ & 0.02 \\
\hline
 QHD & 7.29 & 10.8 & 2.93 & 522 & 540 & --- & --- & --- \\
\hline
\end{tabular}
\end{center}
\end{table}
We note that the nuclear compressibility is higher than that in QMC-I 
($K \sim$ 200 -- 300 MeV)~\cite{sai5}.  However, it is still much 
lower than in QHD~\cite{walecka}.  As in QMC-I, the bag radius of the 
nucleon shrinks a little, while its rms radius swells a little.  On the 
other hand, because of the scalar field, the eigenvalue is reduced more 
than 10\% (at $\rho_0$) from that in free space.  

\begin{figure}[hbt]
\begin{center}
\epsfig{file=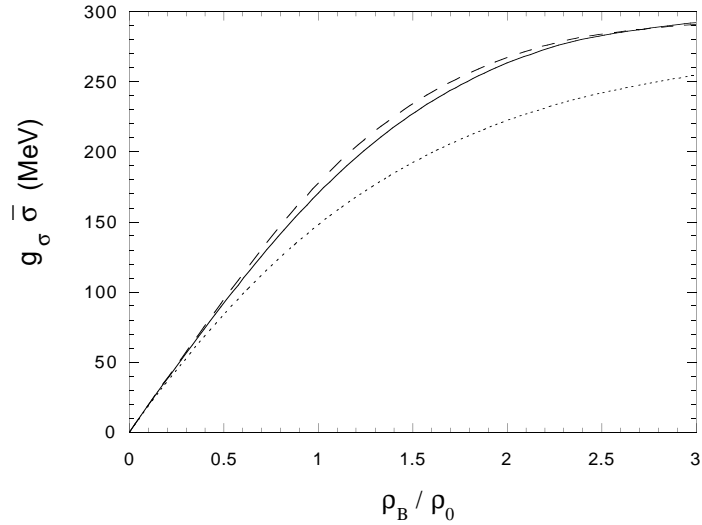,height=7cm}
\caption{Scalar mean-field values.  
The dotted, solid and dashed curves are, respectively, for type A, B 
and C -- as discussed below Eq.(21).}
\label{fig:gssig}
\end{center}
\end{figure}
The strength of the scalar mean-field, $g_{\sigma}{\overline \sigma}$, 
in medium is shown in Fig.\ref{fig:gssig}. 
At small density it is well approximated by a linear function of the density:
\bge
g_{\sigma}{\overline \sigma} \approx 200 \mbox{ (MeV) } 
\left( \frac{\rho_B}{\rho_0} \right). 
\label{appv}
\ene
\clearpage

\subsection{A new scaling phenomenon for hadron masses in matter}
\label{subsec:scaling}

\begin{figure}[hbt]
\begin{center}
\epsfig{file=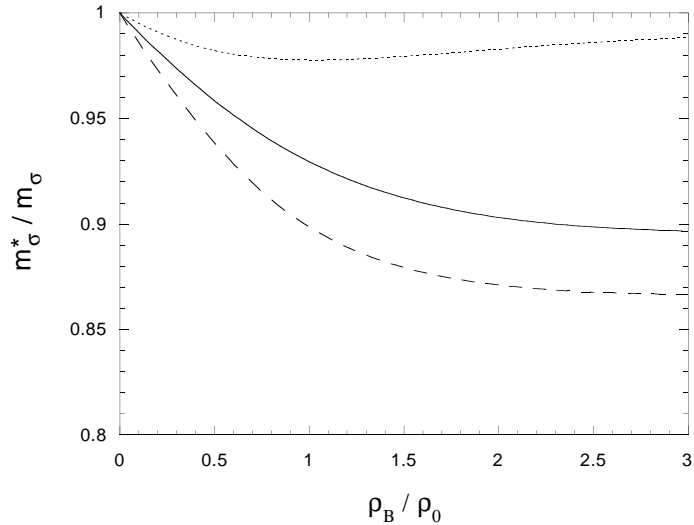,height=7cm}
\caption{Effective $\sigma$-meson mass in symmetric nuclear matter.  
The curves are labelled as in Fig.1.}
\label{fig:esm}
\end{center}
\end{figure}

First, we show the dependence of the $\sigma$-meson mass on the nuclear 
density in Fig.\ref{fig:esm}. 
Using Eqs.(\ref{sigmas}) and (\ref{appv}), we find the $\sigma$ mass at low 
density is 
\bge
\left( \frac{m_\sigma^{\star}}{m_\sigma} \right) \simeq 1 - \alpha_\sigma 
\left( \frac{\rho_B}{\rho_0} \right) , 
\label{sigmas2}
\ene
where $\alpha_\sigma$ = (0.06, 0.1, 0.15) for parameter set
(A, B, C), respectively. 

\begin{figure}[hbt]
\begin{center}
\epsfig{file=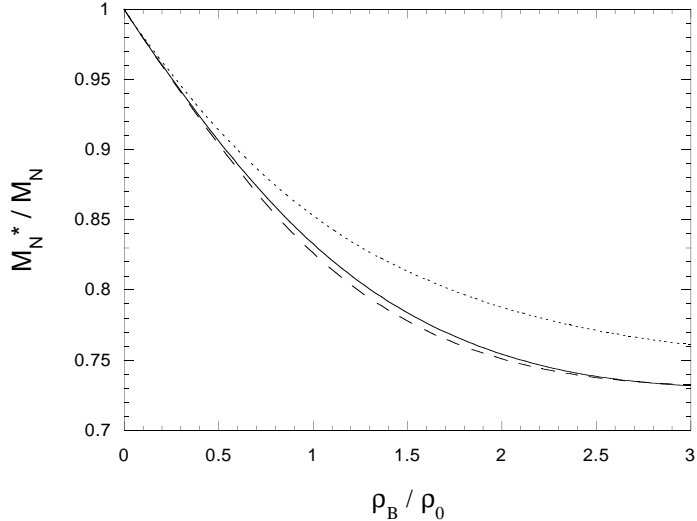,height=7cm}
\caption{Effective nucleon mass in symmetric nuclear matter.  
The curves are labelled as in Fig.1.}
\label{fig:enm}
\end{center}
\end{figure}

The effective nucleon mass is shown in Fig.\ref{fig:enm}.  It decreases 
as the density goes up, and behaves like a constant at large density.  
At small density it is approximately given by using Eqs.(\ref{mstaR}) and 
(\ref{appv}): 
\bge
\left( \frac{M_N^{\star}}{M_N} \right) \simeq 1 - 0.21 
\left( \frac{\rho_B}{\rho_0} \right) . 
\label{nstr2}
\ene
\begin{figure}[ht]
\begin{center}
\epsfig{file=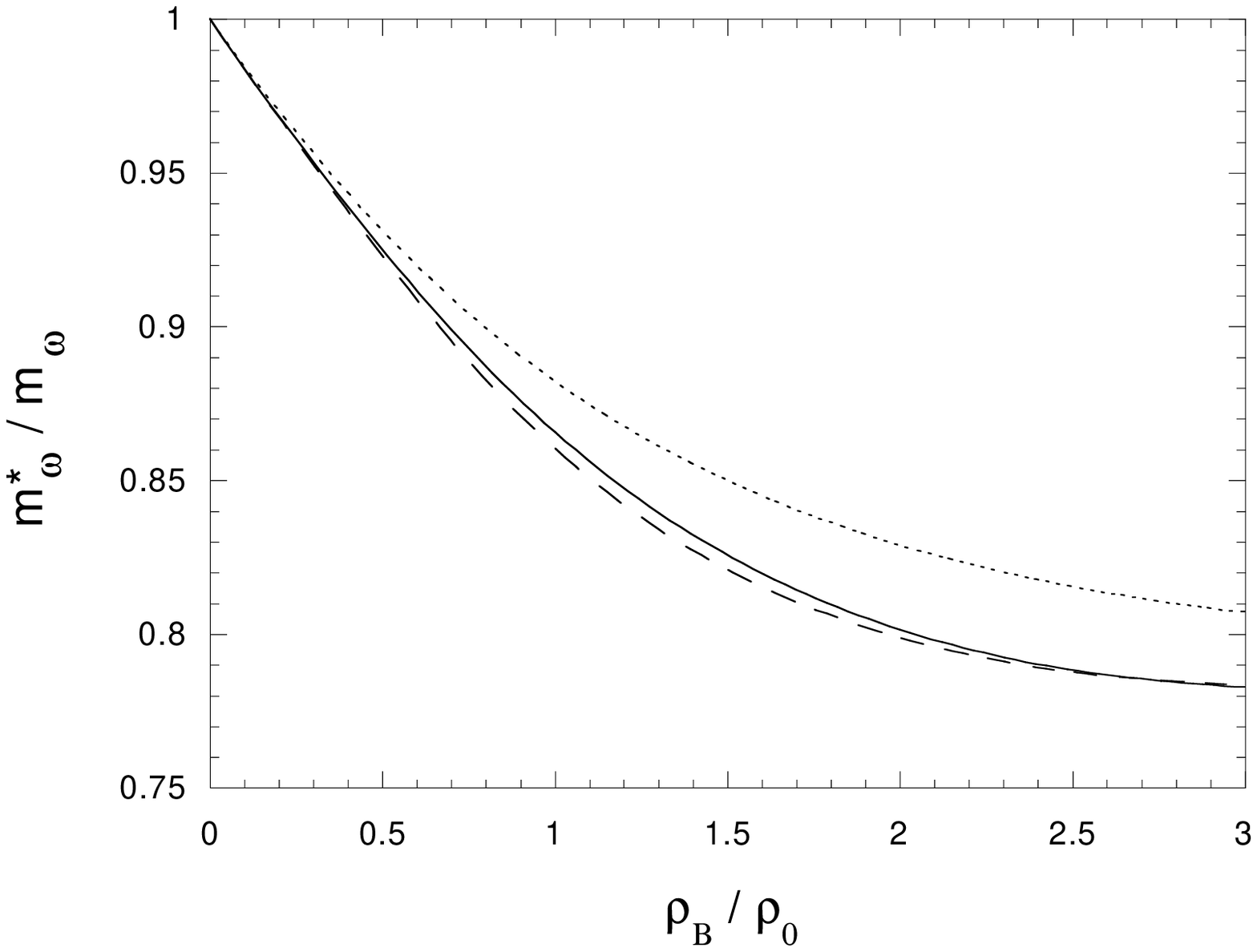,height=7cm}
\caption{Effective ($\rho$- or) $\omega$-meson mass in symmetric nuclear 
matter.  The curves are labelled as in Fig.1.}
\label{fig:eom}
\end{center}
\end{figure}
In Fig.\ref{fig:eom} the effective $\omega$-meson mass is shown as a 
function of the density.  
(Since the difference between the effective $\omega$- and 
$\rho$-meson masses at the same density is very small, we show only 
one curve for both mesons in the figure.)  As the density increases 
the vector-meson mass decreases (as several authors have previously 
noticed~\cite{asa,hatas,sai1,suzuki,will,hat,hat0}) and seems to become 
flat like the effective nucleon mass.  Again, using Eqs.(\ref{vmm}) 
and (\ref{appv}), the mass reduction can be well approximated by a linear 
form at small density:
\bge
\left( \frac{m_v^{\star}}{m_v} \right) \simeq 1 - 
0.17 \left( \frac{\rho_B}{\rho_0} \right). 
\label{vmm2}
\ene
The reduction factor, 0.17, is consistent with other models which have
been applied to the same problem~\cite{hat0}.  

\begin{figure}[ht]
\begin{center}
\epsfig{file=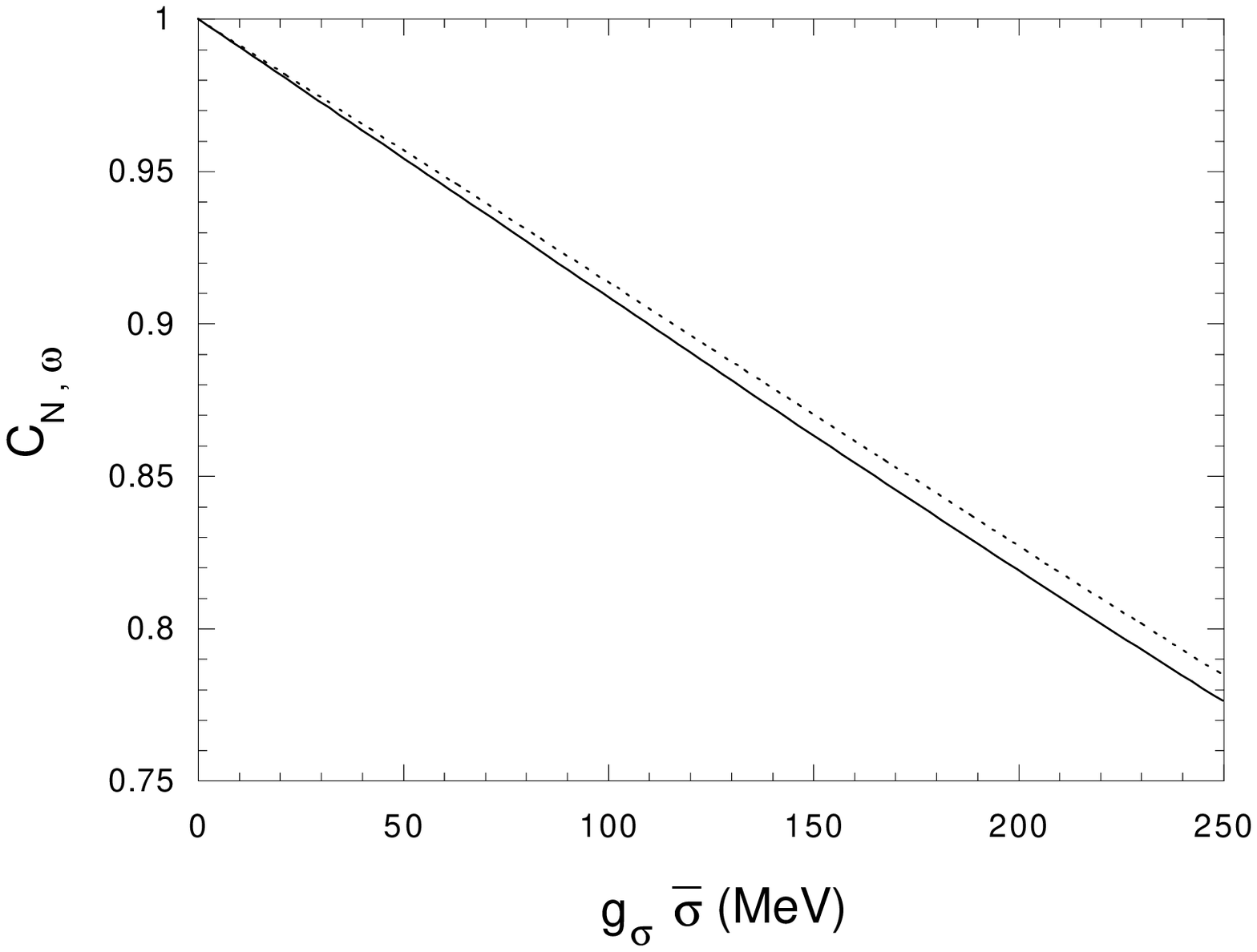,height=7cm}
\caption{The ratios of the quark-scalar density in medium to that in free 
space for the nucleon (solid curve) and the $\omega$ meson (dotted curve)  
-- using parameter set B. }
\label{fig:cno}
\end{center}
\end{figure}
In Fig.\ref{fig:cno} we show the ratios of the quark-scalar density in medium 
to that in free space for the nucleon ($C_N$) and the $\omega$ meson 
($C_\omega$).  As pointed out previously, we can easily see that the ratio for 
the nucleon is well approximated by a linear function of $g_\sigma \sigma$. 
It is also true that the ratio for the vector meson can be  
well described by a similar, linear function of $g_\sigma \sigma$: 
\bge
C_v(\sigma) = 1 - a_v \times (g_{\sigma} \sigma) . 
\label{paramCV}
\ene
We will see this parametrization again later.  

\begin{figure}[hbt]
\begin{center}
\epsfig{file=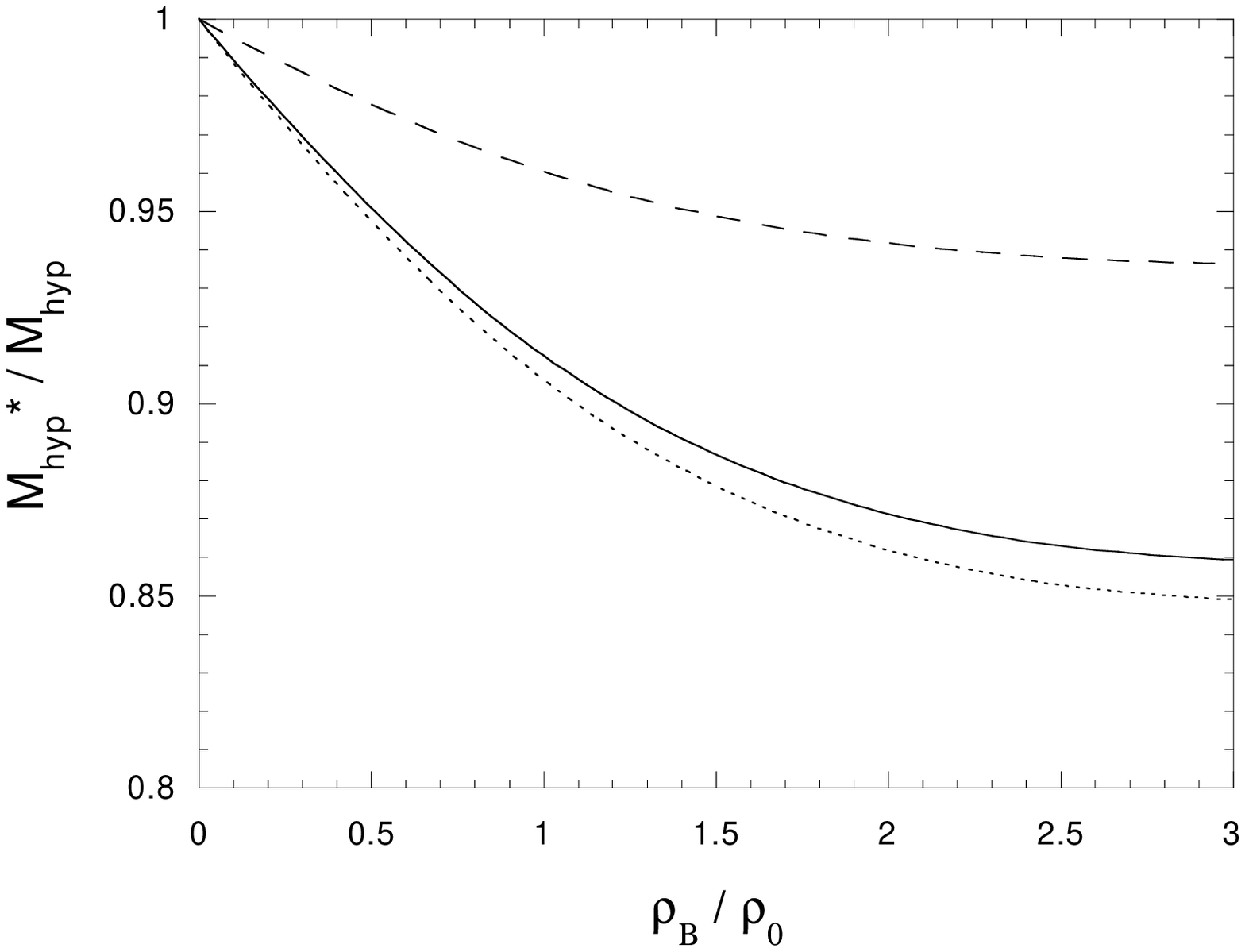,height=7cm}
\caption{The ratio of the hyperon mass in medium to that in free space.  
The dotted, solid and dashed curves are respectively for the $\Lambda$, 
$\Sigma$, and $\Xi$ hyperons, using parameter set B. }
\label{fig:ehm}
\end{center}
\end{figure}

In the present model it is possible to calculate masses of other hadrons.  
In particular, there is considerable interest in studying the masses 
of hyperons in medium -- eg. $\Lambda$, $\Sigma$ and $\Xi$.  
For the hyperons themselves we again use the MIT bag model.  
We assume that the strange quark in the hyperon does not directly couple 
to the scalar field in MFA, as one would expect if the $\sigma$-meson
represented a two-pion-exchange potential. It is also assumed
that the addition of a single hyperon to nuclear matter of density 
$\rho_B$ does not alter the values of the scalar and vector mean-fields, 
namely, we take the local-density approximation to the hyperons~\cite{hyp}.  
The mass of the strange quark, $m_s$, is taken to be $m_s$ = 250 MeV, and 
new $z$-parameters in the mass formula are again introduced to 
reproduce the free hyperon masses: $z_\Lambda$ = 3.131, $z_\Sigma$ = 2.810, 
and $z_\Xi$ = 2.860.  
Using those parameters, we have calculated the 
masses of $\Lambda$, $\Sigma$ and $\Xi$ in symmetric nuclear matter.  They 
are presented in Fig.\ref{fig:ehm}. 
As for the nucleon and the vector mesons, the effective 
mass of the hyperon is determined by only the scalar field.  

In general, we thus find that the effective hadron mass in medium is given by
\bg
M_j^{\star} &=& M_j + \left( \frac{\partial M_j^{\star}}{\partial \sigma} 
\right)_{\sigma=0} \sigma + \frac{1}{2} \left( \frac{\partial^2 M_j^{\star}}
{\partial \sigma^2} \right)_{\sigma=0} \sigma^2 + \cdots , \nn \\
  &\simeq& M_j - \frac{n_0}{3} g_\sigma \Gamma_{j/N} \sigma 
  - \frac{n_0}{6} g_\sigma \Gamma_{j/N} C_j^\prime(0) \sigma^2 , 
\label{hm}
\en
where $j$ stands for $N$, $\omega$, $\rho$, $\Lambda$, $\Sigma$, $\Xi$, etc., 
$n_0$ is the number of non-strange quarks in the hadron $j$, 
$\Gamma_{j/N} = S_j(0)/S_N(0)$ with the quark-scalar density, $S_j$, in 
$j$, and the scalar density ratio, $C_j(\sigma) = S_j(\sigma)/S_j(0)$.  

Using Eqs.(\ref{appv}) and (\ref{hm}), we find that the hyperon masses at 
low density are given by 
\bge
\left( \frac{M_\Lambda^{\star}}{M_\Lambda} \right) \simeq 1 - 
0.12 \left( \frac{\rho_B}{\rho_0} \right) , 
\label{lambm}
\ene
\bge
\left( \frac{M_\Sigma^{\star}}{M_\Sigma} \right) \simeq 1 - 
0.11 \left( \frac{\rho_B}{\rho_0} \right) , 
\label{sigmam}
\ene
and
\bge
\left( \frac{M_\Xi^{\star}}{M_\Xi} \right) \simeq 1 - 
0.05 \left( \frac{\rho_B}{\rho_0} \right) , 
\label{lambm2}
\ene
where we take $\Gamma_{\Lambda,\Sigma,\Xi/N} =1$, because we find that 
the $\Gamma$ factor for the hyperon is again quite close to unity (e.g. 
$\Gamma_{\Lambda/N} = 1.0001$, in our actual calculations). 

As seen in Fig.\ref{fig:cno} the linear approximation to the scalar-density 
ratio, $C_j$, is very convenient.  We find that it is numerically 
relevant to not only the nucleon and the vector mesons but also the 
hyperons: 
\bge
C_j(\sigma) = 1 - a_j \times (g_{\sigma} \sigma) , 
\label{paramH}
\ene
where $a_j$ is the slope parameter for the hadron $j$.  We list them in 
Table~\ref{slope}. 
\begin{table}[htbp]
\begin{center}
\caption{Slope parameters for the hadrons ($\times 10^{-4}$ 
MeV$^{-1}$). }
\label{slope}
\begin{tabular}[t]{c|cccccc}
\hline
type & $a_N$ & $a_\omega$ & $a_\rho$ & $a_\Lambda$ & $a_\Sigma$ & 
$a_\Xi$ \\
\hline
 A &  9.01 & 8.63 & 8.59 & 9.27 & 9.52 & 9.41 \\
 B &  8.98 & 8.63 & 8.58 & 9.29 & 9.53 & 9.43 \\
 C &  8.97 & 8.63 & 8.58 & 9.29 & 9.53 & 9.43 \\
\hline
\end{tabular}
\end{center}
\end{table}
We should note that the dependence of $a_j$ on the hadrons is quite weak, 
and it ranges around $8.6 \sim 9.5 \times 10^{-4}$ (MeV$^{-1}$).  

If we ignore the weak dependence of $a_j$ on the hadrons and take 
$\Gamma_{j/N}=1$ in Eq.(\ref{hm}), the effective hadron mass can be 
rewritten in a quite simple form: 
\bge
M_j^{\star} \simeq M_j - \frac{n_0}{3} (g_\sigma \sigma) \left[ 1 - 
  \frac{a}{2} (g_\sigma \sigma) \right] , 
\label{hm3}
\ene
where $a \simeq 9.0 \times 10^{-4}$ (MeV$^{-1}$).  This mass formula can 
reproduce the hadron masses in 
matter quite well over a wide range of $\rho_B$, up to $\sim 3 \rho_0$.

Since the scalar field is common to all hadrons, Eq.(\ref{hm3}) leads to 
a new, simple scaling relationship among the hadron masses: 
\bge
\left( \frac{\delta m_v^{\star}}{\delta M_N^{\star}} \right) \simeq 
\left( \frac{\delta M_\Lambda^{\star}}{\delta M_N^{\star}} \right) \simeq 
\left( \frac{\delta M_\Sigma^{\star}}{\delta M_N^{\star}} \right) \simeq 
\frac{2}{3} \ \ \ \mbox{ and } \ \ \ 
\left( \frac{\delta M_\Xi^{\star}}{\delta M_N^{\star}} \right) \simeq 
\frac{1}{3} , 
\label{scale}
\ene
where $\delta M_j^{\star} \equiv M_j - M_j^{\star}$.  The factors, 
$\frac{2}{3}$ 
and $\frac{1}{3}$, in Eq.(\ref{scale}) come from the ratio of the number of 
non-strange quarks in $j$ to that in the nucleon.  This means that the hadron 
mass is practically determined by only the number of non-strange quarks, 
which feel the common scalar field generated by surrounding nucleons in medium, and the strength of the scalar field~\cite{sai4}.  On the other hand, 
the change in the confinement mechanism due to the environment gives a small 
contribution to the above ratio.  
It would be very interesting to see whether this scaling relationship is 
correct in forthcoming experiments.  
\clearpage

\subsection{Finite nuclei}
\label{subsec:finite}

In this subsection we will show our results for some finite, closed shell 
nuclei. 
The Lagrangian density, Eq.(\ref{qmc-2}), leads to the 
following equations for finite nuclei: 
\bg
\frac{d^2}{dr^2} \sigma(r) + \frac{2}{r} \frac{d}{dr} \sigma(r) 
    - m_\sigma^{\star 2} \sigma(r) &=& - g_\sigma C_N \rho_s(r) 
    - m_\sigma m_\sigma^{\star} g_\sigma [ a_\sigma - 2 b_\sigma g_\sigma 
      \sigma(r) ] \sigma(r)^2 \nn \\ 
    &+& \frac{2}{3} g_\sigma [ m_\omega^{\star} \Gamma_{\omega/N} 
      C_\omega \omega(r)^2 
    + m_\rho^{\star} \Gamma_{\rho/N} C_\rho b(r)^2 ] , \\
\label{scmot}
\frac{d^2}{dr^2} \omega(r) + \frac{2}{r} \frac{d}{dr} \omega(r) 
    - m_\omega^{\star 2} \omega(r) &=& - g_\omega \rho_B(r) , \\
\label{vcmot}
\frac{d^2}{dr^2} b(r) + \frac{2}{r} \frac{d}{dr} b(r) 
    - m_\rho^{\star 2} b(r) &=& - \frac{g_\rho}{2} \rho_3(r) , \\
\label{rhmot}
\frac{d^2}{dr^2} A(r) + \frac{2}{r} \frac{d}{dr} A(r) 
    &=& - e \rho_p(r) , 
\label{gammot}
\en
where 
\bg
\rho_s(r) &=&  \sum_\alpha^{occ} d_\alpha(r)
    (|G_\alpha(r)|^2 - |F_\alpha(r)|^2), \label{scrden} \\
\rho_B(r) &=&  \sum_\alpha^{occ} d_\alpha(r)
    (|G_\alpha(r)|^2 + |F_\alpha(r)|^2), \label{baryden} \\
\rho_3(r) &=&  \sum_\alpha^{occ} 
    d_\alpha(r) (-)^{t_\alpha -1/2} 
    (|G_\alpha(r)|^2 + |F_\alpha(r)|^2), \label{rho3} \\
\rho_p(r) &=& \sum_\alpha^{occ} d_\alpha(r) 
    (t_\alpha + \frac{1}{2}) 
    (|G_\alpha(r)|^2 + |F_\alpha(r)|^2), \label{phtn3} 
\en
with $d_\alpha(r)= (2j_\alpha+1)/4\pi r^2$, and 
\bg
\frac{d}{dr} G_\alpha(r) + \frac{\kappa}{r} G_\alpha(r) - 
\left[ \epsilon_\alpha - g_\omega \omega(r) - t_\alpha g_\rho b(r)
\right. 
&-& \left. (t_\alpha + \frac{1}{2}) e A(r) + M_N \right. \nn \\
&-& \left. g_\sigma(\sigma(r)) \sigma(r) \right] F_\alpha(r) = 0 , 
\label{qwave1} \\
\frac{d}{dr} F_\alpha(r) - \frac{\kappa}{r} F_\alpha(r) + 
\left[ \epsilon_\alpha - g_\omega \omega(r) - t_\alpha g_\rho b(r)
\right.
&-& \left. (t_\alpha + \frac{1}{2}) e A(r) - M_N \right. \nn \\
&+& \left. g_\sigma (\sigma(r)) \sigma (r) \right] G_\alpha(r) = 0 . 
\label{qwave2} 
\en
Here $G_\alpha(r)/r$ and $F_\alpha(r)/r$ are respectively the 
radial part of the upper and lower 
components of the solution to the Dirac equation for the nucleon: 
\bge
\psi({\vec r}) = {i[G_\alpha(r)/r] \Phi_{\kappa m} \choose  
-[F_\alpha(r)/r] \Phi_{-\kappa m}} \xi_{t_\alpha}, 
\label{wave}
\ene
where $\xi_{t_\alpha}$ is a two-component isospinor and $\Phi_{\kappa m}$ is 
a spin spherical harmonic~\cite{horowitz} 
($\alpha$ labelling the quantum numbers and $\epsilon_\alpha$ being 
the energy).  Then, the normalization condition is 
\bge
\int dr (|G_\alpha(r)|^2 + |F_\alpha(r)|^2) =1 . \label{norm}
\ene
As usual, $\kappa$ specifies the angular quantum numbers and $t_\alpha$ 
the eigenvalue of the isospin operator $\tau^N_3/2$.  
Practically, $m_\sigma^{\star}$, $m_v^{\star}$ and $C_j$ are 
respectively given by Eqs.(\ref{sigmas}), (\ref{hm}) and (\ref{paramH}), and 
$g_\sigma(\sigma(r))$ is 
\bge
g_\sigma (\sigma(r)) =  g_\sigma \left[ 1 - \frac{a_N}{2} 
            (g_\sigma \sigma(r)) \right] . 
\label{ccg}
\ene
The total energy of the system is then given by 
\bg
E_{tot} &=& \sum_\alpha^{occ} (2j_\alpha + 1) \epsilon_\alpha 
 - \frac{1}{2} \int d{\vec r} \ [ -g_\sigma D(\sigma (r)) \sigma(r) \nn \\
 &+& g_\omega \omega(r) \rho_B(r) + \frac{1}{2} g_\rho b(r) \rho_3(r) 
 + eA(r) \rho_p(r) ] ,  \label{ftoten}
\en
where
\bg
D(\sigma(r)) &=& C_N \rho_s(r) 
    + m_\sigma m_\sigma^{\star} [ a_\sigma - 2 b_\sigma g_\sigma 
      \sigma(r) ] \sigma(r)^2 \nn \\ 
    &-& \frac{2}{3} \left[ m_\omega^{\star} \Gamma_{\omega/N} 
      C_\omega \omega(r)^2 + m_\rho^{\star} \Gamma_{\rho/N} C_\rho 
    b(r)^2 \right] . 
\label{dd}
\en

There are seven parameters to be determined: $g_\sigma$, $g_\omega$, 
$g_\rho$, $e$, $m_\sigma$, $m_\omega$ and $m_\rho$.  As in 
the case of infinite matter we take the 
experimental values: $m_\omega$ = 783 MeV, $m_\rho$ = 770 MeV and 
$e^2/4\pi$ = 1/137.036.  
The coupling constants $g_\sigma$, $g_\omega$ and $g_\rho$ are fixed to 
describe the nuclear matter properties and 
the bulk symmetry energy per baryon of 35 MeV (see Table~\ref{ccc}).  

\begin{table}[htbp]
\begin{center}
\caption{Model parameters for finite nuclei (for $m_q$ = 5 MeV and 
$R_N$ = 0.8 fm). }
\label{ccc2}
\begin{tabular}[t]{c|cccc}
\hline
 type & $g_{\sigma}^2/4\pi$ & $g_{\omega}^2/4\pi$ & 
$g_{\rho}^2/4\pi$ & $m_\sigma$(MeV) \\
\hline
 A & 1.67 & 2.70 & 5.54 & 363 \\
\hline
 B & 2.01 & 3.17 & 5.27 & 393 \\
\hline
 C & 2.19 & 3.31 & 5.18 & 416 \\
\hline
\end{tabular}
\end{center}
\end{table}
The $\sigma$-meson mass however determines the range of the attractive 
interaction and changes in $m_\sigma$ affect the nuclear-surface slope and 
its thickness.  Therefore, as in the paper of Horowitz and 
Serot~\cite{horowitz}, we adjust $m_\sigma$ to fit the measured rms charge 
radius of $^{40}$Ca, $r_{ch}$($^{40}$Ca) = 3.48 fm~\cite{sinha}. 
(Notice that variations of $m_\sigma$ at fixed 
($g_\sigma / m_\sigma$) have no effect on the infinite nuclear matter 
properties~\cite{sai5}.)  
We summarize the parameters in Table~\ref{ccc2}.  

Equations (\ref{scmot}) to (\ref{norm}) give a set of coupled non-linear 
differential equations, which may be solved by a standard iteration 
procedure~\cite{cal}.

\begin{figure}[htb]
\centering{\
\epsfig{file=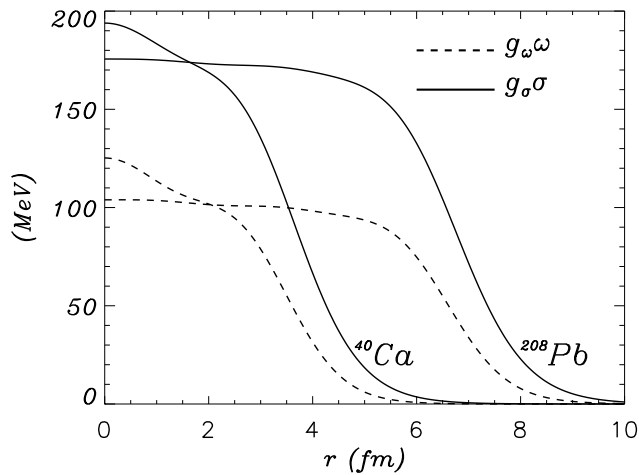,height=7cm}
\caption{Scalar and vector strength for $^{40}$Ca and $^{208}$Pb 
(for type B). }
\label{fsso}}
\end{figure}
In Fig.\ref{fsso} we first show the calculated strength of the $\sigma$ 
and $\omega$ fields in $^{40}$Ca and $^{208}$Pb.  
\begin{figure}[htb]
\centering{\
\epsfig{file=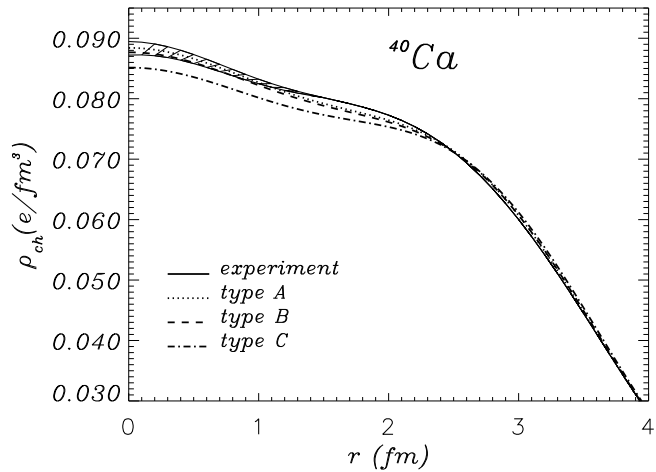,height=7cm}
\caption{Charge density distribution for $^{40}$Ca compared with the 
experimental data. 
}
\label{chca40}}
\end{figure}
\begin{figure}[htb]
\centering{\
\epsfig{file=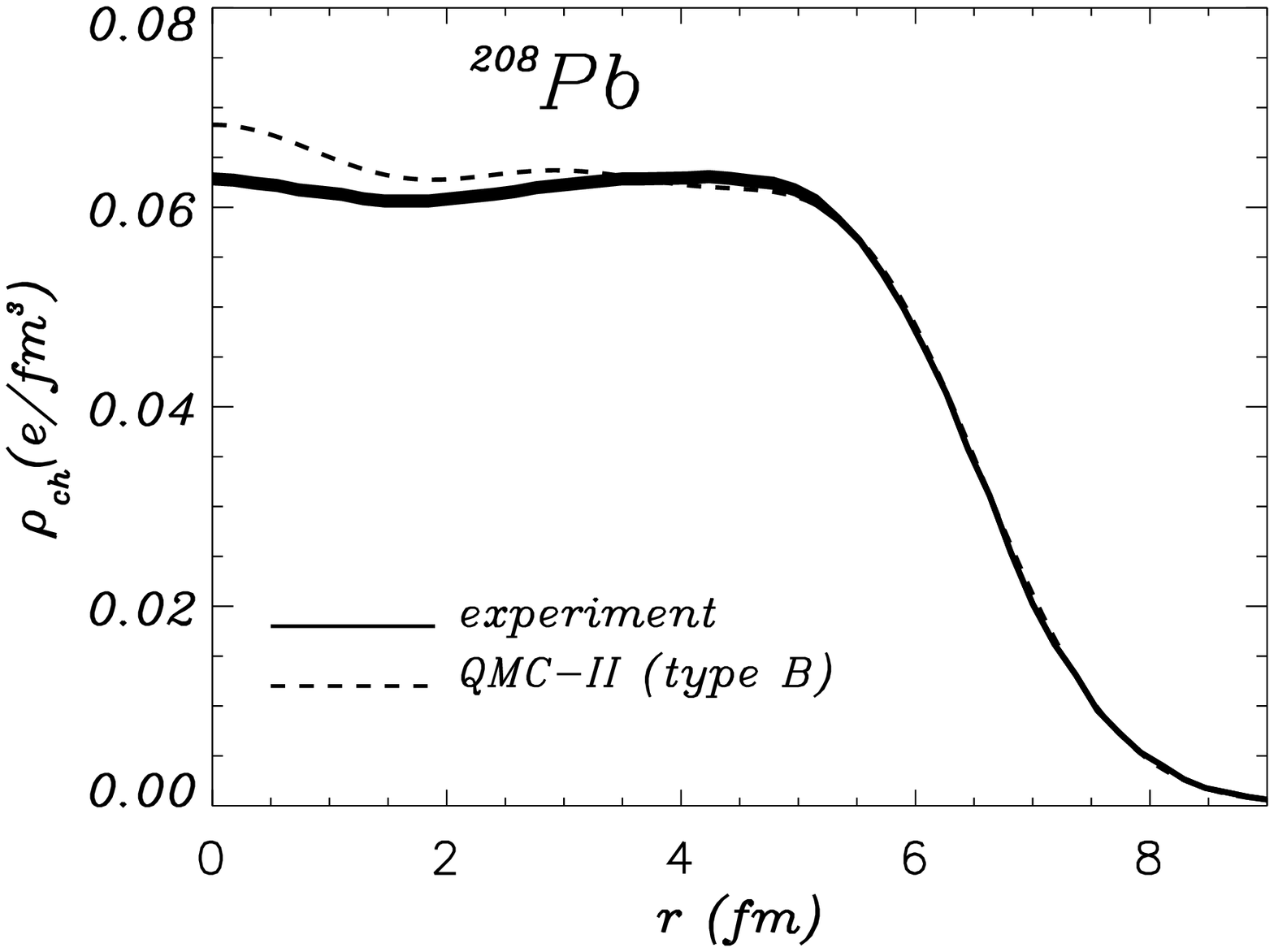,height=7cm}
\caption{Same as Fig.8 but for $^{208}$Pb.  The parameter set B is used. }
\label{chpb}}
\end{figure}
Next we show calculated charge density distributions, $\rho_{ch}$, 
of $^{40}$Ca and $^{208}$Pb in comparison with those of the experimental 
data in Figs.\ref{chca40} and \ref{chpb}.  To see the difference among the 
results from the three parametrizations of $m_\sigma^{\star}$ (A, B and C), 
in Fig.\ref{chca40} we present only the interior part of 
$\rho_{ch}(^{40}$Ca).  
As in Ref.\cite{sai5}, we have used a convolution of the point-proton 
density, which is given by solving Eqs.(\ref{scmot}) $\sim$ (\ref{norm}), 
with the proton charge distribution to calculate $\rho_{ch}$.  For $^{40}$Ca 
the QMC-II model with parameter sets A and B give similar charge 
distributions to those in QMC-I, while the result of  QMC-II with  
parameter set C is closer to that in QHD.  From 
Fig.\ref{chpb} we see that the present model also yields a  
charge distribution for $^{208}$Pb which is similar to those calculated using 
QMC-I or QHD.  

\begin{table}[htbp]
\begin{center}
\caption{Calculated proton and neutron spectra of $^{40}$Ca (for type B) 
compared with QMC-I and the experimental data ($m_q$ = 5 MeV and 
$R_N$ = 0.8 fm).  Here, I and II denote, respectively, QMC-I and 
QMC-II.  All energies are in MeV. }
\label{spca40}
\begin{tabular}[t]{c|cccccc}
\hline
 & \multicolumn{3}{c}{neutron} & 
\multicolumn{3}{c}{proton} \\
\cline{2-7} 
Shell & I & II & Expt. & I & II & Expt. \\
\hline
$1s_{1/2}$ & 43.1 & 41.1 & 51.9 & 35.2 & 33.2 & 50$\pm$10 \\
$1p_{3/2}$ & 31.4 & 30.0 & 36.6 & 23.8 & 22.3 & 34$\pm$6 \\
$1p_{1/2}$ & 30.2 & 29.0 & 34.5 & 22.5 & 21.4 & 34$\pm$6 \\
$1d_{5/2}$ & 19.1 & 18.0 & 21.6 & 11.7 & 10.6 & 15.5 \\
$2s_{1/2}$ & 15.8 & 14.7 & 18.9 &  8.5 &  7.4 & 10.9 \\
$1d_{3/2}$ & 17.0 & 16.4 & 18.4 &  9.7 &  9.0 & 8.3  \\
\hline
\end{tabular}
\end{center}
\end{table}
In Table~\ref{spca40}, the calculated spectrum of $^{40}$Ca is presented.
Because of the relatively smaller scalar and vector fields in the present 
model than in QHD, the spin-orbit splittings are smaller (in this 
respect the model is very similar to QMC-I). 
We should note that there is a strong correlation between the 
effective nucleon mass and the spin-orbit force~\cite{sai5}.  
The problem concerning the spin-orbit force in the QMC model has been
studied in 
Refs.\cite{guichon1,sai5,blun,jin1}. It remains to be seen whether the
higher order corrections, as studied by Wallace et al.~\cite{wallace},
will help to resolve it.

Table~\ref{sum} gives a summary of the calculated binding energy per nucleon 
($E/A$), rms charge radii and the difference between nuclear rms radii for 
neutrons and protons ($r_n - r_p$), for several closed-shell nuclei.  
\begin{table}[htbp]
\begin{center}
\caption{Binding energy per nucleon, $- E/A$ (in MeV), rms charge radius 
$r_{ch}$ (in fm) and the difference between $r_n$ and $r_p$ (in fm) for type B, 
$m_q$ = 5 MeV and $R_B$ = 0.8 fm. 
I and II denote, respectively, QMC-I and QMC-II. ($^*$ fit) }
\label{sum}
\begin{tabular}[t]{c|ccc|ccc|ccc}
\hline
 & & $-E/A$ & & & $r_{ch}$ & & & $r_n-r_p$ & \\
\hline
model&I&II&Expt.&I&II&Expt.&I&II& Expt. \\
\hline
$^{16}$O &5.84&5.11&7.98&2.79&2.77&2.73&$-0.03$&$-0.03$&0.0 \\
$^{40}$Ca&7.36&6.54&8.45&3.48$^*$&3.48$^*$&3.48&$-0.05$&$-0.05$&0.05$\pm$0.05\\
$^{48}$Ca&7.26&6.27&8.57&3.52&3.53&3.47&0.23&0.24&0.2$\pm$0.05 \\
$^{90}$Zr&7.79&6.99&8.66&4.27&4.28&4.27&0.11&0.12&0.05$\pm$0.1 \\
$^{208}$Pb&7.25&6.52&7.86&5.49&5.49&5.50&0.26&0.27&0.16$\pm$0.05 \\
\hline
\end{tabular}
\end{center}
\end{table}
While there are still some discrepancies between 
the results and data, the present model provides reasonable results.  
In particular, as in QMC-I, it reproduces the rms charge radii for medium 
and heavy nuclei quite well.  

\begin{figure}[htb]
\centering{\
\epsfig{file=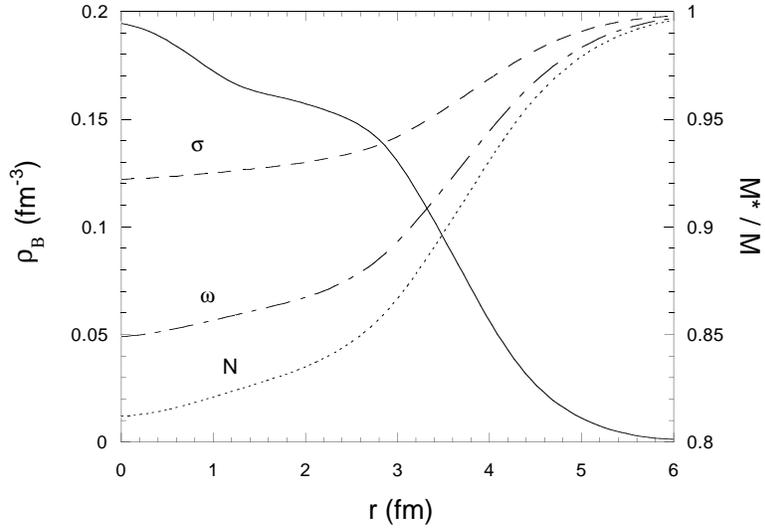,height=7cm}
\caption{Changes of the nucleon, $\sigma$ and $\omega$ meson masses 
in $^{40}$Ca.  The nuclear baryon density is also illustrated (solid curve).  
The right (left) scale is for the effective mass (the baryon density). 
The parameter set B is used.  
}
\label{hmca40}}
\end{figure}
\begin{figure}[htb]
\centering{\
\epsfig{file=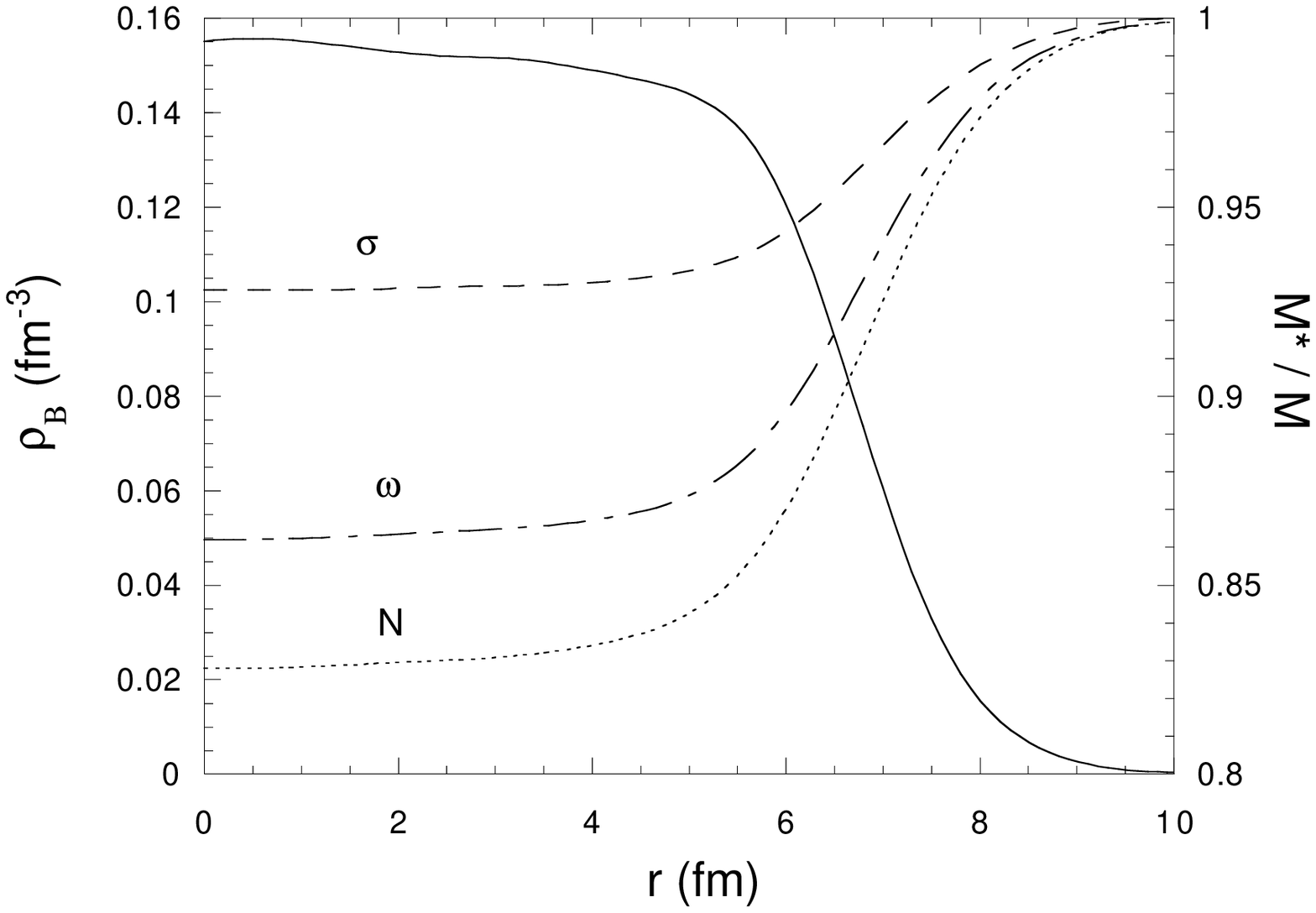,height=7cm}
\caption{Same as Fig.10 but for $^{208}$Pb. 
}
\label{hmpb}}
\end{figure}
In Figs.\ref{hmca40} and \ref{hmpb} we present the changes of the 
nucleon, $\sigma$ and $\omega$ meson masses in $^{40}$Ca and $^{208}$Pb, 
respectively.  The interior density of $^{40}$Ca is much higher than 
$\rho_0$, while that in $^{208}$Pb is quite close to $\rho_0$.  Accordingly, 
in the interior the effective hadron masses in $^{40}$Ca 
are smaller than in $^{208}$Pb.  We can also see this in  
Fig.\ref{fsso}, 
where the strength of the scalar field in the interior part of $^{40}$Ca is 
stronger than in $^{208}$Pb.  

\begin{figure}[htb]
\centering{\
\epsfig{file=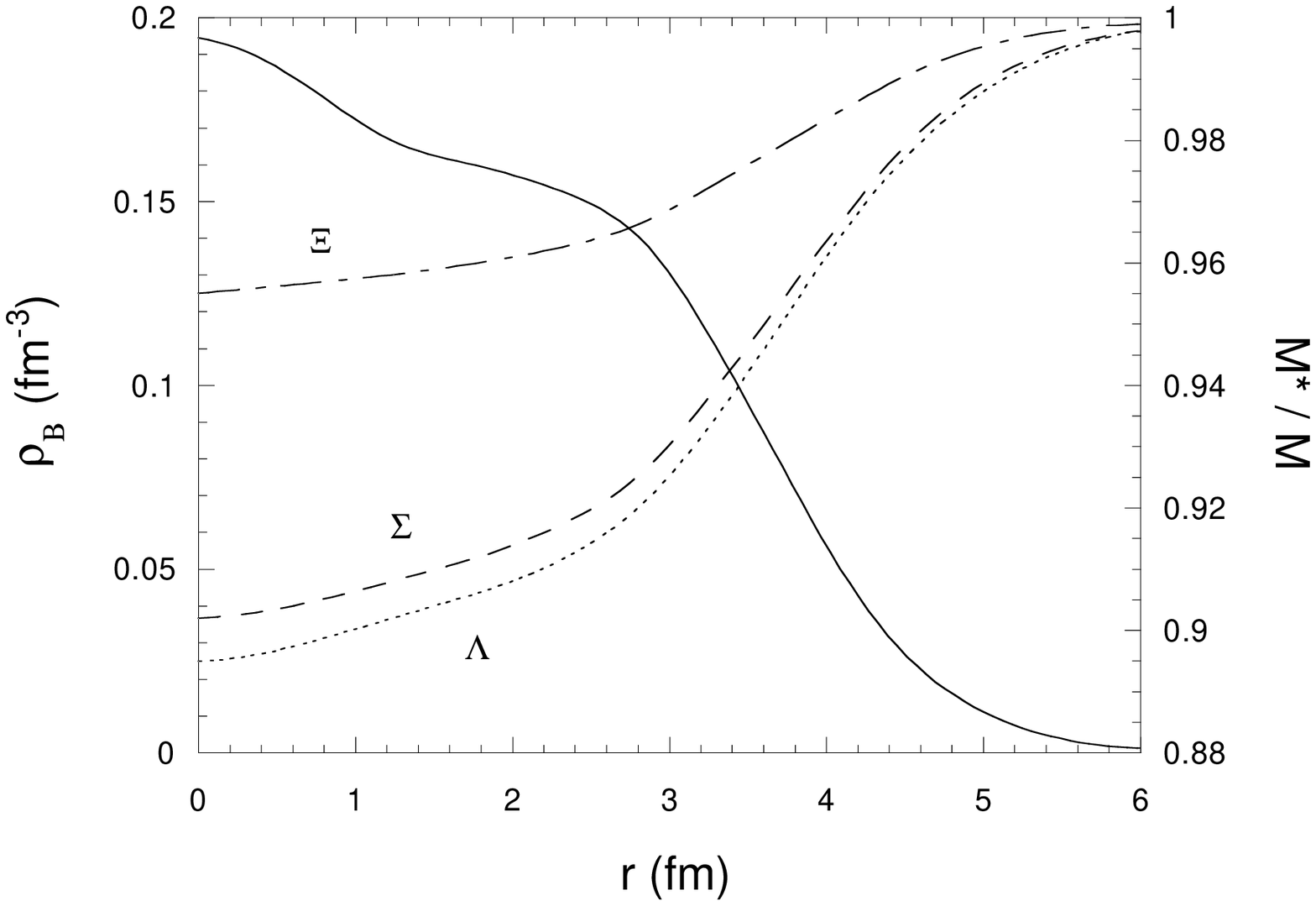,height=7cm}
\caption{Changes of the hyperon ($\Lambda$, $\Sigma$ and $\Xi$) masses 
in $^{40}$Ca.  The solid curve is for the nuclear baryon density. 
The right (left) scale is for the effective mass (the baryon density). 
The parameter set B is used. }
\label{hypca40}}
\end{figure}
\begin{figure}[htb]
\centering{\
\epsfig{file=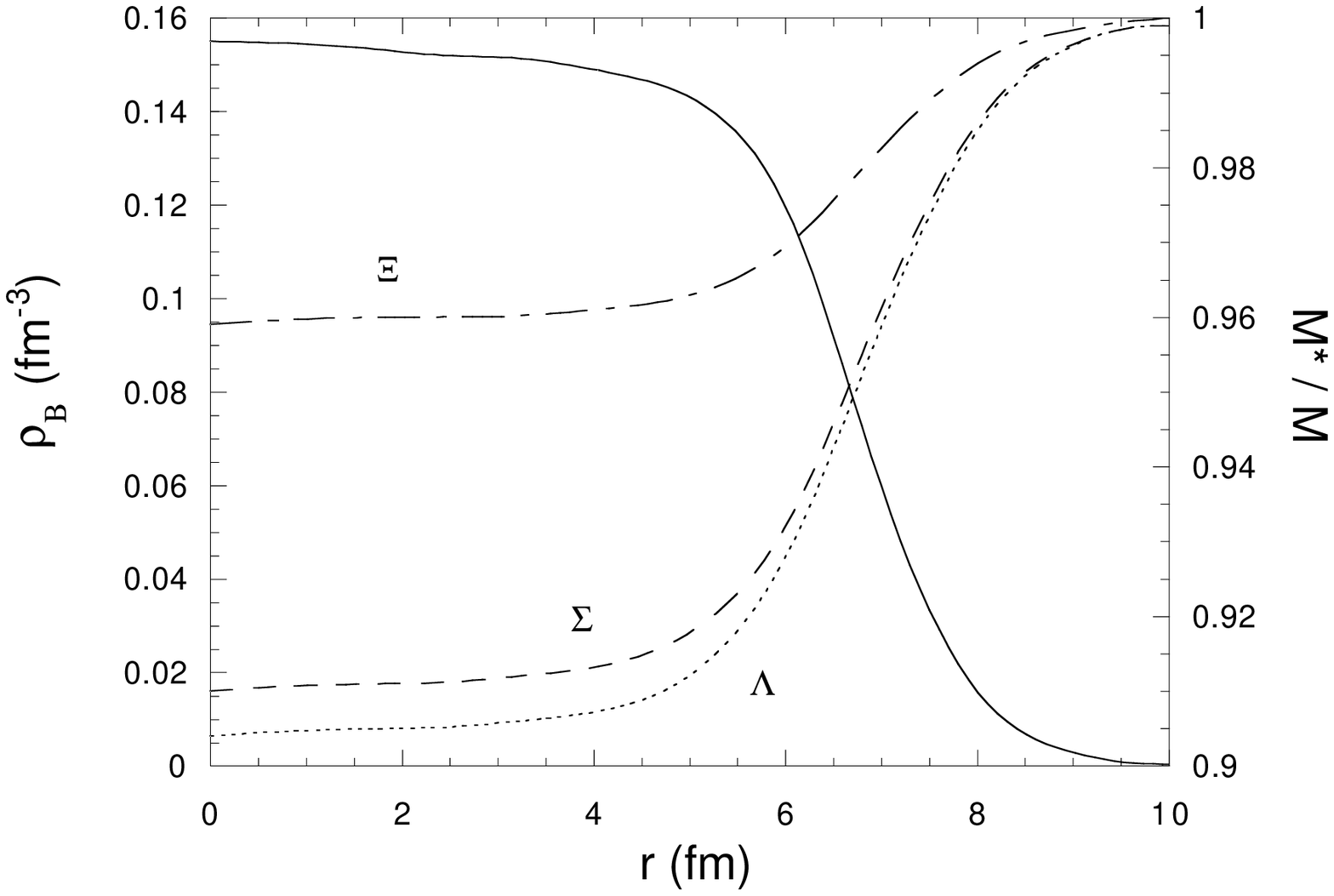,height=7cm}
\caption{Same as Fig.12 but for $^{208}$Pb. 
}
\label{hyppb}}
\end{figure}
Using the local-density approximation and Eq.(\ref{hm3}), 
it is possible to 
calculate the changes of the hyperon ($\Lambda$, $\Sigma$ and $\Xi$) masses 
in $^{40}$Ca and $^{208}$Pb, which are respectively illustrated in 
Figs.\ref{hypca40} and \ref{hyppb}.  Our quantitative calculations for 
the changes of the hyperon masses in finite nuclei may be quite important 
in forthcoming experiments concerning hypernuclei~\cite{hyp}.  

\clearpage

\section{CONCLUSION}
\label{sec:concl}

We have extended the quark-meson coupling (QMC) model to include  
quark degrees of freedom within the scalar and vector mesons, as well as 
in the nucleons, and have investigated 
the density dependence of hadron masses in nuclear medium.  
As several authors have 
suggested~\cite{asa,hatas,sai1,suzuki,will,hat,hat0,sai4}, the hadron mass is 
reduced because of the scalar mean-field in medium.  Our results 
are quite consistent with the other models. 
In the present model the hadron mass can be related to the number of 
non-strange quarks and the strength of the scalar mean-field 
(see Eq.(\ref{hm3})).  
We have found a new, simple formula to describe the hadron masses in the 
medium, and this led to a new scaling relationship among them 
(see Eq.(\ref{scale})).  
Furthermore, we have calculated the changes of not only the nucleon, 
$\sigma$, $\omega$ and $\rho$ masses but also the hyperon 
($\Lambda$, $\Sigma$ and $\Xi$) masses in finite nuclei.  
We should note that the origins of the mass reduction in QMC and QHD are 
completely different~\cite{sai2}.  
It would be very interesting to compare our results with forthcoming 
experiments on hypernuclei.  

By applying this extended QMC model to finite nuclei, we have studied the 
properties of some static, closed shell nuclei.  
Our (self-consistent) calculations reproduce well the observed 
static properties of nuclei such as the charge density distributions.  
In the present model, there are, however, still some discrepancies in 
energy spectra of nuclei, in particular, the spin-orbit splittings.  
To overcome this defect, we have discussed one possible way, in which a 
constituent quark mass ($\sim 300$ MeV) is adopted, in 
Refs.\cite{guichon1,sai5}.  As an alternative, Jin and 
Jennings~\cite{jin1} and Blunden and Miller~\cite{blun} have proposed  
variations of the bag constant and $z$ parameter in medium, which 
have been suggested by the fact that quarks are partially deconfined in 
matter.  To help settle this problem, one should perhaps 
consider the change of the vacuum 
properties in the medium~\cite{hatkun}.  

Our Lagrangian density, Eq.(\ref{qmc-2}), provides a lot of effective 
coupling terms among the meson fields because the mesons have structure (cf. 
Ref.\cite{serot2}).  In particular, the Lagrangian 
automatically offers self-coupling terms (or non-linear terms) with 
respect to the $\sigma$ field.  Using Eq.(\ref{sigmas}), the Lagrangian 
density gives the non-linear $\sigma$ terms (up to ${\cal O}(\sigma^4)$) as: 
\bg
{\cal L}_{QMC-II}^{NL\sigma} &=& - \frac{1}{2} m_\sigma^{\star}(\sigma)^2 
\sigma^2,  \nn \\
&\simeq& -\frac{1}{2}m_\sigma^2 \sigma^2 + g_\sigma a_\sigma m_\sigma^2 
\sigma^3 - \frac{1}{2} g_\sigma^2 (a_\sigma^2 + 2 b_\sigma) m_\sigma^2 
\sigma^4 . 
\label{nls}
\en

On the other hand, in nuclear physics, QHD with non-linear $\sigma$ terms 
has been extensively used in MFA to describe realistic nuclei~\cite{ring0}. 
The most popular parametrizations are called NL1, NL2~\cite{rein} and 
NL-SH~\cite{ring}, and the non-linear terms in those parametrizations 
are given as 
\bge
{\cal L}_{QHD}^{NL\sigma} = - \frac{1}{2} m_\sigma^2 \sigma^2 + 
\frac{1}{3} g_2 \sigma^3 + \frac{1}{4} g_3 \sigma^4 , 
\label{nls2}
\ene
where $g_2$ and $g_3$ take respectively a positive 
(negative) [positive] and positive (negative) [positive] values  
in NL1 (NL2) [NL-SH].  Since the non-linear $\sigma$ terms provide the 
self-energy of $\sigma$ meson, it changes the $\sigma$ mass in matter.  
Comparing Eq.(\ref{nls2}) with Eq.(\ref{nls}), we can see that 
the effective $\sigma$ mass in NL2 {\em increases\/} at low nuclear 
density while the $\sigma$ mass {\em decreases\/} in NL1 and NL-SH in MFA.  

However, from the point of view of a field theory, like the 
Nambu--Jona-Lasinio model, an increase of the $\sigma$ mass in the 
medium seems unlikely~\cite{hatkun,bern}.  (We should note 
that the values of $g_2$ in those parametrizations are small compared 
with the corresponding one in Eq.(\ref{nls}).)  
Furthermore, from the point of view of field theory, 
$g_3$ in Eq.(\ref{nls2}) should be negative because the vacuum must be 
stable~\cite{lee}.  Therefore, we can conclude that 
one would expect to find $g_2 \geq 0$ and $g_3 \leq 0$ 
in Eq.(\ref{nls2}).  Unfortunately, the above three parametrizations used in 
nuclear physics do not satisfy the condition, while our Lagrangian, 
Eq.(\ref{nls}), does.  
It will be very interesting to explore the connection 
between various coupling strengths found empirically in earlier work and 
those found in our approach.  

Finally, we would like to give some caveats concerning the present 
calculation.  The basic idea of the model is that the mesons are 
locally coupled to the quarks.  Therefore, in the present model the 
effect of short-range correlations among the quarks, which would be associated
with overlap of the hadrons, are completely neglected.  
At very high density these would be expected to dominate and the present 
model must eventually break down there (probably beyond $\sim 
3\rho_B/\rho_0$).  Furthermore, the pionic cloud of the 
hadron~\cite{cloudy} should be considered explicitly in any truly 
quantitative study of hadron properties in medium.  
We note that subtleties such as scalar-vector mixing in medium and the 
splitting between longitudinal and transverse masses of the vector 
mesons~\cite{will} have been ignored in the present mean-field study. 
Although the former appears to be quite
small in QHD the latter will certainly be important in any attempt to
actually measure the mass shift.

\vspace{0.5cm}
This work was supported by the Australian Research Council.
%
%

%


\begin{thebibliography}{99}
\bibitem{cmcor} In our earlier works, the 
phenomenological center-of-mass correction to the bag energy was used.  
In this paper, however, we do not use it because the c.m. and gluon 
fluctuation corrections may be absorbed into the familiar form, 
$-z_N/R_N$.  See Ref.\cite{guichon1}.
\end{thebibliography}
\end{document}